\documentclass[twocolumn]{aastex631}

\submitjournal{ApJ Letters}
\shorttitle{Anisotropic Satellite Quenching as SMBH Feedback Signature}
\shortauthors{Karp, Lange and Wechsler}

\newcommand{\hmpc}{h^{-1} \, \mathrm{Mpc}}
\newcommand{\hmsun}{h^{-1} \, M_\odot}
\newcommand{\mpeak}{M_{\rm peak}}
\newcommand{\aacc}{a_{\rm acc}}

\begin{document}

\title{Anisotropic Satellite Galaxy Quenching: A Unique Signature of Energetic Feedback by Supermassive Black Holes?}

\correspondingauthor{Juliana Karp}
\email{juliana.karp@yale.edu}

\author[0000-0002-1728-8042]{Juliana S. M. Karp}
\affil{Kavli Institute for Particle Astrophysics and Cosmology, Stanford University, CA 94305, USA}
\affil{Department of Astronomy, Yale University, New Haven, CT 06511, USA}

\author[0000-0002-2450-1366]{Johannes U. Lange}
\affil{Kavli Institute for Particle Astrophysics and Cosmology, Stanford University, CA 94305, USA}
\affil{Department of Astronomy and Astrophysics, University of California, Santa Cruz, CA 95064, USA}
\affil{Department of Physics, University of Michigan, Ann Arbor, MI 48109, USA}
\affil{Leinweber Center for Theoretical Physics, University of Michigan, Ann Arbor, MI 48109, USA}

\author[0000-0003-2229-011X]{Risa H. Wechsler}
\affil{Kavli Institute for Particle Astrophysics and Cosmology, Stanford University, CA 94305, USA}
\affil{Department of Physics, Stanford University, CA 94305, USA}
\affil{SLAC National Accelerator Laboratory, Menlo Park, CA 94025, USA}

\begin{abstract}
    The quenched fraction of satellite galaxies is aligned with the orientation of the halo’s central galaxy, such that on average, satellites form stars at a lower rate along the major axis of the central. This effect, called anisotropic satellite galaxy quenching (ASGQ), has been found in observational data and cosmological simulations. Analyzing the IllustrisTNG simulation, \cite{MartinNavarro2021_Natur_594_187} recently argued that ASGQ is caused by anisotropic energetic feedback and constitutes ``compelling observational evidence for the role of black holes in regulating galaxy evolution.'' In this letter, we study the causes of ASGQ in state-of-the-art galaxy formation simulations to evaluate this claim. We show that cosmological simulations predict that on average, satellite galaxies along the major axis of the dark matter halo tend to have been accreted at earlier cosmic times and are hosted by subhalos of larger peak halo masses. As a result, a modulation of the quenched fraction with respect to the major axis of the central galaxy is a natural prediction of hierarchical structure formation. We show that ASGQ is predicted by the UniverseMachine galaxy formation model, a model without anisotropic feedback. Furthermore, we demonstrate that even in the IllustrisTNG simulation, anisotropic satellite accretion properties are the main cause of ASGQ. Ultimately, we argue that ASGQ is not a reliable indicator of supermassive black hole feedback in galaxy formation simulations and, thus, should not be interpreted as such in observational data.
\end{abstract}


\section{Introduction}

Galaxies reside within dark matter halos that form via hierarchical accretion. They can be classified into central galaxies --- those residing at the halo centers --- and accreted satellite galaxies --- those which orbit the halo potential. The mechanisms that affect the star formation rate (SFR) of galaxies and the times at which they stop forming stars (i.e., become quenched and red) are active areas of research. It is commonly assumed that satellite galaxies quench at a faster rate due to satellite-specific effects in the high-density environment of the halo, such as ram pressure stripping, galaxy interactions, and the cutoff of gas supply \citep[see][for a review]{Mo2010_gfe_book}. For massive central galaxies, most theoretical models of galaxy formation postulate that interactions with supermassive black holes (SMBHs) regulate their star formation. Given the importance of SMBH feedback in contemporary models of galaxy formation, observational constraints on the proposed SMBH-galaxy interactions are critical.

Large galaxy surveys such as the \textit{Sloan Digital Sky Survey} (SDSS) have shed light on the galaxy populations of galaxy groups and clusters. It is now well-established that the distribution of satellite galaxies within halos is spatially aligned with the orientation of their central galaxies \citep{Brainerd2005_ApJ_628_101}. The strength of this alignment depends on satellite galaxy properties, such as stellar mass $M_\star$, SFR, and color. \citet{Yang2006_MNRAS_369_1293} found that red satellite galaxies, i.e., those with lower SFRs, in galaxy groups and clusters are more strongly spatially aligned with the major axis of the central galaxy than their blue, star-forming counterparts. This finding can be re-formulated into an effect called \textit{Anisotropic Satellite Galaxy Quenching} (ASGQ): satellite galaxies along the major axis of the central galaxy are redder and form fewer stars than those along the minor axis of the central.

Recently, \citet[][hereafter MN21]{MartinNavarro2021_Natur_594_187} argued that the observed ASGQ effect in SDSS is evidence for energetic feedback by SMBHs residing in the central galaxies. The authors show that the ASGQ effect is also observed in the IllustrisTNG hydro-dynamical simulation of galaxy formation \citep{Nelson2019_ComAC_6_2}. MN21 suggest that ASGQ in IllustrisTNG is caused by SMBHs residing in central galaxies, and that the observed ASGQ effect in SDSS galaxy groups and clusters can be attributed to the same source. MN21 show that outflows driven by the SMBHs \citep{Weinberger2017_MNRAS_465_3291} are aligned with the minor axis of the centrals and lower the density of the circumgalactic medium (CGM) along the minor axis of the central. The scenario proposed by MN21 is that due to the reduced CGM density, quenching by ram pressure stripping is reduced for satellites passing along the central galaxy's minor axis, giving rise to the observed ASGQ effect. We will refer to this hypothesis as the anisotropic feedback scenario.

However, anisotropic feedback is not the only possible explanation for ASGQ. \citet[][hereafter K07]{Kang2007_MNRAS_378_1531} were able to qualitatively and quantitatively reproduce the ASGQ effect observed in SDSS in semi-analytic models (SAMs) of galaxy formation. SAMs typically do not model galaxy orientation or three-dimensional outflows. Thus, the mechanism proposed by MN21 cannot explain the findings in K07. Instead, K07 argue that satellite galaxies accreted along the major axis of the dark matter halo were hosted by more massive halos when they were accreted. Given the anti-correlation between satellite galaxy SFR and halo mass at accretion, it is expected that satellites along the major axis of the halo form fewer stars on average. Finally, the preferential alignment of central galaxies, typically ellipticals, in groups and clusters \citep{Brown2002_MNRAS_333_501, Brainerd2005_ApJ_628_101, Georgiou2021_AA_647_185} then gives rise to ASGQ without the need for an anisotropic quenching mechanism acting on satellite galaxies. This second mechanism relies on anisotropic satellite accretion properties to explain ASGQ. Overall, the findings of K07 caution against the interpretation of ASGQ as direct evidence for SMBH-galaxy interactions.

In this letter, we evaluate the proposed mechanisms for the ASGQ effect observed in SDSS by studying ASGQ in the UniverseMachine \citep{Behroozi2019_MNRAS_488_3143} galaxy formation model and re-examining MN21's findings in the IllustrisTNG simulations.

\section{Simulations}

\begin{figure*}
    \centering
    \includegraphics{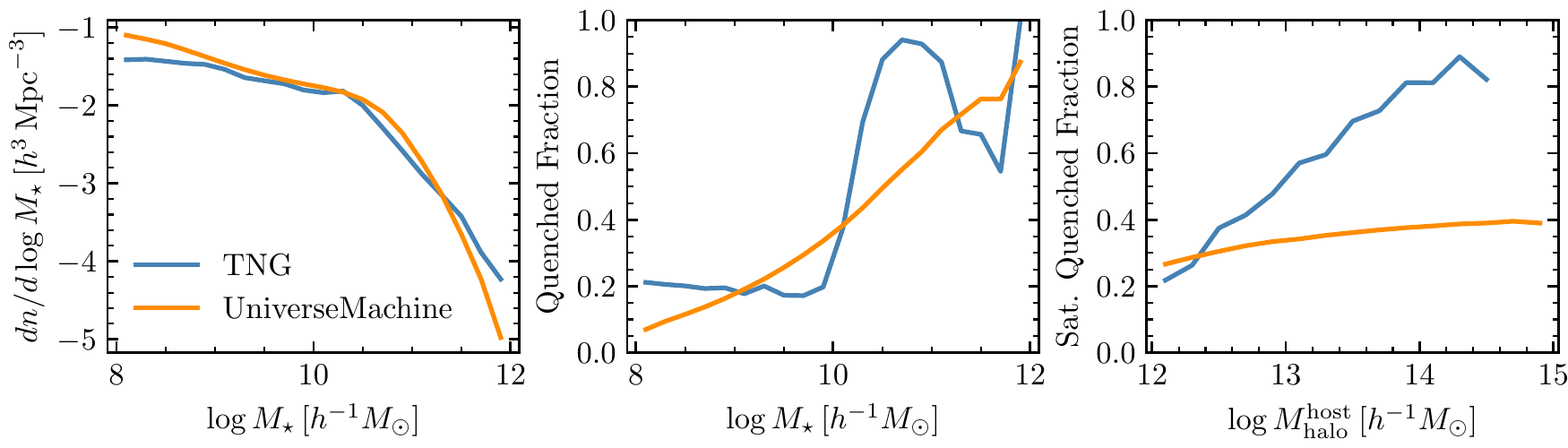}
    \caption{The stellar mass function (left), stellar mass-dependent galaxy quenched fraction (middle) and host halo mass-dependent satellite quenched fraction (right) of galaxies in TNG100 (blue) and UniverseMachine run on SMDPL (orange).}
    \label{fig:um_vs_tng}
\end{figure*}

 In this letter, we compare results from the $z = 0$ snapshot of the TNG100 simulation and the UniverseMachine galaxy formation model run on the Small Multidark Planck (SMDPL) simulation. Both simulations assume a $\Lambda$CDM cosmology consistent with results from \citet{PlanckCollaboration2016_AA_594_13}. TNG100 is part of the IllustrisTNG suite of hydrodynamical galaxy formation simulations \citep{Nelson2019_ComAC_6_2}. It has a cubic box size of $75 \, \hmpc$ in side length, where $h$ is the dimensionless Hubble parameter. TNG100 contains $1820^3$ initial gas cells, $1820^3$ dark matter particles, and $2 \times 1820^3$ Monte Carlo tracer particles for baryons. The dark matter and target baryon mass resolutions are $5.1 \times 10^6 \, \hmsun$ and $9.4 \times 10^5 \, \hmsun$, respectively. SMDPL is part of the MultiDark cosmological dark matter-only simulation suite \citep{Klypin2016_MNRAS_457_4340}. It has a cubic box size of $400 \, \hmpc$ in side length and contains $3840^3$ particles with a mass resolution of $9.6 \times 10^7 \, \hmsun$. Halo catalogs were produced using the {\sc ROCKSTAR} halo finder \citep{Behroozi2013_ApJ_762_109}. UniverseMachine \citep{Behroozi2019_MNRAS_488_3143} is an empirical post-processing algorithm that models galaxy formation by relating the SFR of galaxies to the accretion rate of their dark matter halos across cosmic time and has been calibrated to reproduce galaxy populations down to $10^7 M_\odot$ in stellar mass.

In both TNG100 and UniverseMachine run on SMDPL, we study satellite galaxies of stellar masses greater than or equal to $10^8 \, \hmsun$ that orbit central galaxies of halo masses greater than $10^{12} \, \hmsun$. These cuts were chosen for consistency with MN21. To define which galaxies are quenched in UniverseMachine, we first establish a star-forming main sequence (SFMS). We separate galaxies into bins of $0.1$ dex in stellar mass and apply a two-component Gaussian mixture model to the $\log \mathrm{sSFR}$ values, where $\mathrm{sSFR} = \mathrm{SFR} / M_\star$ is the specific SFR. We take the peak of the high-sSFR Gaussian component as the location of the SFMS. Following the definition in MN21, we define galaxies as quenched if their sSFR is more than $1.0$ dex below the SFMS. To define galaxies in TNG100 as quenched, we follow the definition used in MN21. Fig.~\ref{fig:um_vs_tng} gives an overview of key galaxy properties in the two simulations.

Finally, for TNG100, we use S\'ersic fits to synthetic galaxy images \citep{RodriguezGomez2019_MNRAS_483_4140} as proxies for the central galaxy orientation in SDSS.

\section{Results}

\begin{figure}[b]
    \centering
    \includegraphics{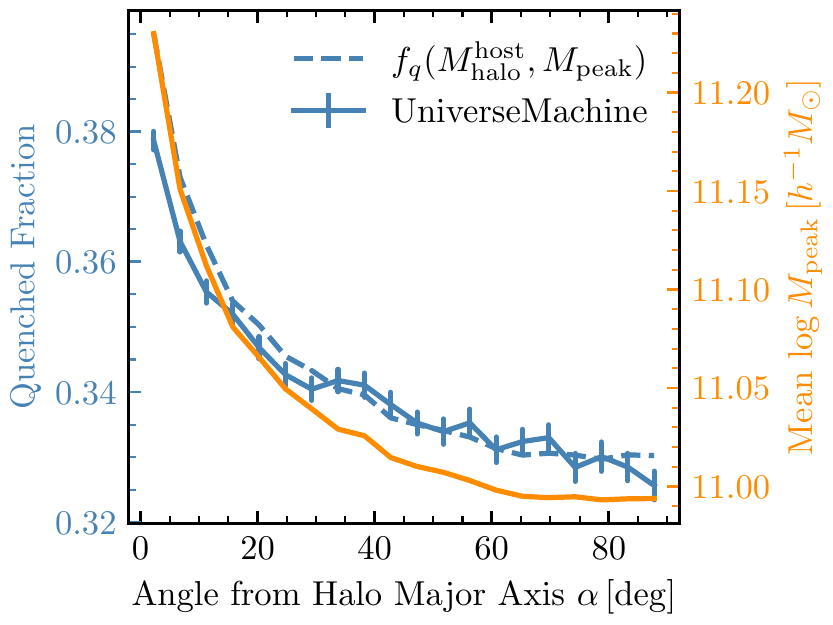}
    \caption{Angular-dependent quenched fraction of satellites (blue error bars), mean peak halo mass (orange), and expected quenched fraction based on central stellar mass and satellite peak halo mass (dashed blue) as a function of angle from major axis of the host halo in UniverseMachine run on SMDPL.}
    \label{fig:um_quenched_mpeak_driver}
\end{figure}

In order to investigate MN21's explanation of the causes of ASGQ --- that ASGQ is a direct result of anisotropic SMBH-driven outflows affecting star formation in satellite galaxies --- we first examine this effect in UniverseMachine run on SMDPL. UniverseMachine is a modern empirical galaxy formation model that, like most SAMs, does not model SMBH-driven three-dimensional outflows. Thus, in UniverseMachine, star-formation feedback on satellite galaxies is modeled isotropically, just like in the SAM analyzed in K07. We compute each satellite galaxy's projected angle $\alpha$ with respect to its host halo's projected major axis, i.e.,
\begin{equation}
    \cos \alpha = \frac{\left| d_x A_x + d_y A_y \right|}{\sqrt{d_x^2 + d_y^2} \sqrt{A_x^2 + A_y^2}} \, ,
\end{equation}
where $\mathbf{d}$ is the difference in the position of the satellite and the central and $\mathbf{A}$ is the direction of halo major axis \citep{Behroozi2013_ApJ_762_109}. Following MN21, we then study the properties of satellite galaxies as a function of their angle from the halo major axis. In the following analysis, we do not exclude galaxies without an associated dark matter subhalo (sometimes referred to as orphan galaxies), but note that our qualitative conclusions would not change if we did. As shown in Figure \ref{fig:um_quenched_mpeak_driver}, we find that satellite galaxies are preferentially quenched along the major axis of their host halos. While galaxy orientations are not part of the UniverseMachine model, central galaxies and their host halos are expected to be directionally aligned \citep[see, e.g.,][]{Brown2002_MNRAS_333_501, Brainerd2005_ApJ_628_101, Georgiou2021_AA_647_185}. Under this assumption, UniverseMachine predicts ASGQ. Because UniverseMachine does not explicitly model interactions between SMBH feedback and the CGM, we infer that a different physical mechanism than that proposed by MN21 must be responsible for ASGQ in this simulation.

What is the cause of the modulation in UniverseMachine? In Figure~\ref{fig:um_quenched_mpeak_driver}, we also show the mean $\log \mpeak$, a subhalo's peak halo mass over its lifetime, as a function of angle from the major axis of the host halo. We find that the trends in quenched fraction and peak halo mass are similar. Considering that galaxies with higher $\mpeak$ are more likely to be quenched, this suggests that ASGQ is driven by differences in the satellite accretion properties based on the alignment of satellites and the dark matter halo shape, in agreement with the scenario proposed by K07. In order to determine whether K07's claimed causal relationship holds true in UniverseMachine, we separate satellites into bins of $0.1$ dex in host halo mass. In each bin, we fit the measured quenched fraction as a function of $\log \mpeak$ with a fourth-degree polynomial to establish an expected quenched fraction based on galaxies' $\log \mpeak$ and host halo mass $M_{\rm halo}^{\rm host}$. Then, using the same stacking method described above, we overplot this expected ASGQ modulation in Figure~\ref{fig:um_quenched_mpeak_driver}. We find that the measured ASGQ is qualitatively consistent with a model where quenching is dependent on $\mpeak$ and $M_{\rm halo}^{\rm host}$, providing strong evidence for the scenario introduced in K07.

\begin{figure}
    \centering
    \includegraphics{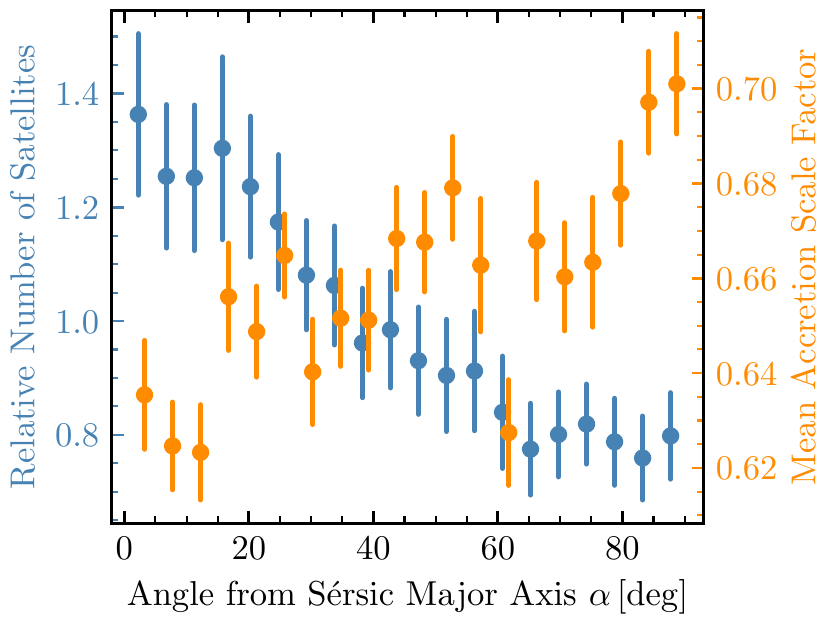}
    \caption{Relative number of satellite galaxies (blue) and mean satellite accretion scale factor (orange) as a function of the projected angle from the major axis of the central galaxy in TNG100.}
    \label{fig:tng_n_a_acc}
\end{figure}

Next, we consider the modulation in TNG100, a modern hydrodynamical galaxy formation simulation analyzed by MN21, which does model SMBH-driven three-dimensional outflows and their interactions with the CGM. TNG100 has a roughly $150$ times smaller volume than UniverseMachine run on SMDPL, making predictions significantly noisier. We estimate $68 \%$ confidence intervals through jackknife re-sampling of $200$ groups of host halos. In Figure~\ref{fig:tng_n_a_acc}, we show that there exists an overdensity of satellite galaxies along the major axis of the central galaxy. This overdensity of satellite galaxies along the major axis of the central implies that central galaxies and their host halos are preferentially aligned in TNG100. We confirm this by estimating halo shapes from the eigenvalues of the halo mass distribution tensor within $r_{\rm 500c}$, similar to \cite{Behroozi2013_ApJ_762_109}. We find that for halos more massive than $10^{12} \ \hmsun$ and $10^{13} \ \hmsun$, the median angular offsets are $29^\circ$ and $16^\circ$, respectively, i.e. below the value of $45^\circ$ expected for a random distribution. In the same figure, we show the distribution of the mean accretion scale factor $a_{\rm acc}$ which we define as the first snapshot a galaxy has been identified as a satellite in the TNG merger trees. We find that, on average, satellite galaxies along the major axis have been accreted earlier than those along the minor axis of the central. We find a similar modulation when calculating the angle with respect to the major axis of the halo\footnote{The anisotropic distribution of $a_{\rm acc}$ depends on the radius out to which the halo mass tensor is calculated. When using all particles within each friends-of-friends (FoF) group to calculate the mass tensor and, ultimately, the halo major axis, we do not find evidence for anisotropy in $a_{\rm acc}$ as a function of the angle from the halo major axis. This indicates that the anisotropy exists primarily with respect to the inner halo structure.}. Thus, we see a modulation in accretion properties along the major axis of the halo and central galaxy, similar to what is found in UniverseMachine.

\begin{figure}
    \includegraphics[width=\columnwidth]{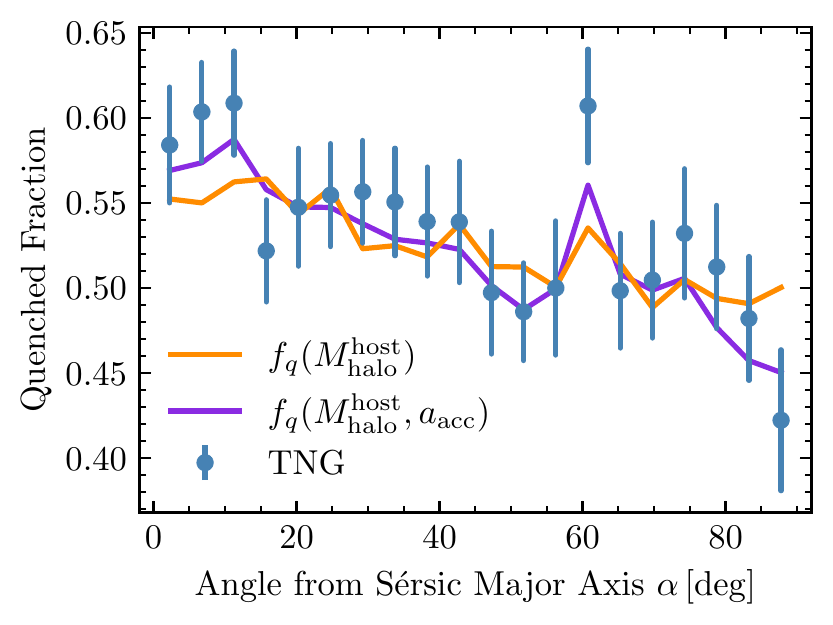}
    \caption{ASGQ signal in TNG100. Blue error bars show the projected ASGQ signal directly measured in TNG100, reproducing the results in MN21. The purple line shows the expected signal from an isotropic quenching model where the quenched fraction only depends on host galaxy stellar mass and satellite accretion scale factor as measured in TNG100. The orange line indicates the expected modulation from the dependence of the satellite quenched fraction on host stellar mass. We find that the ASGQ effect in TNG100 can be explained without the need for anisotropic feedback originating from SMBH-driven outflows affecting star formation in satellite galaxies.}
    \label{fig:tng_quenched_drivers_2}
\end{figure}

We then investigate whether the ASGQ modulation in TNG100 can be entirely explained by these accretion properties without the need for explicit anisotropic feedback induced by interactions between SMBH outflows and the CGM. To quantify the amplitude of ASGQ, we fit the quenched fraction of satellite galaxies versus angle from the central major axis with a linear relationship using generalized least squares. Analyzing the ASGQ modulation in TNG100 shown in Figure~\ref{fig:tng_quenched_drivers_2}, we find a slope of $(-13.6 \pm 2.7) \times 10^{-4} \, \mathrm{deg}^{-1}$. We find similar results by deriving uncertainties in the measured slope value from Monte-Carlo realizations of TNG100 with random central galaxy orientations. In order to investigate the causes of ASGQ in TNG, we employ a similar method as in the analysis of UniverseMachine. We divide satellites into $0.25$ dex wide bins in host halo mass $M_{\rm halo}^{\rm host}$. Using a fourth-degree polynomial, we fit the measured quenched fraction in each bin as a function of the satellites' accretion scale factor $\aacc$. We thus establish an expected quenched fraction $f_q(M_{\rm halo}^{\rm host}, \aacc)$ based on galaxies' $\aacc$ and host halo mass $M_{\rm halo}^{\rm host}$. This predicted quenched fraction, overplotted in Figure~\ref{fig:tng_quenched_drivers_2}, exhibits a modulation of slope $(-13.2 \pm 2.0) \times 10^{-4} \, \mathrm{deg}^{-1}$. The residual amplitude, obtained by subtracting the expected from the measured modulation, is $(-0.9 \pm 1.7) \times 10^{-4} \, \mathrm{deg}^{-1}$, i.e., consistent with zero. In other words, the ASGQ amplitude in TNG100 can be entirely explained by the accretion properties of satellites without the need for angular-dependent quenching originating from three-dimensional SMBH-driven outflows.

We note that, unlike in UniverseMachine, the total ASGQ amplitude is better predicted by the dependence of satellite quenching on $\aacc$ rather than $\mpeak$. In fact, we detect with only $\sim 1 \sigma$ confidence that $\mpeak$ decreases along the major axis of the central, whereas the significance is $\sim 8\sigma$ for $\aacc$ increasing along the major axis. Finally, we point out that the stacking procedure can artificially cause part of the observed modulation. The quenched fraction of satellites may vary with host halo properties, such as total halo mass, as shown in the right panel of Fig.~\ref{fig:um_vs_tng}. Since the alignment between centrals and halos also increases with halo mass, this can cause a modulation in the quenched fraction in the stack, even if there is no modulation for individual halos. Indeed, we find a $\sim 4 \sigma$ significant angular dependence in the host halo mass of satellites. The expected ASGQ amplitude solely based on this effect is also shown in Figure~\ref{fig:tng_quenched_drivers_2} and accounts for an amplitude of $(-5.8 \pm 1.5) \times 10^{-4} \, \mathrm{deg}^{-1}$, i.e., roughly half of the observed effect.

\section{Discussion}

Our results indicate that the anisotropic satellite quenching (ASGQ) phenomenon is a natural consequence of hierarchical cosmic structure formation. Central galaxies are preferentially aligned with the shapes of dark matter halos which, in turn, tend to accrete matter along filaments of the cosmic web. Subhalos found along the major axis of the halo, on average, have larger peak halo masses and entered the halo at earlier cosmic times. These trends are likely at least partially caused by anisotropic accretion along cosmic filaments \citep{Wang2005_MNRAS_364_424}. Additionally, they may also be the result of internal processes whereby the alignment between subhalo orbits after infall and the halo shape may depend on subhalo properties. Because satellite quenching is positively correlated with accretion mass and time spent in the host halo \citep[e.g.][and references therein]{Wechsler2018_ARAA_56_435}, this leads to the expectation that the quenched fraction of satellite galaxies is lower along the central's minor axis. Thus, ASGQ can be produced without invoking SMBH-driven outflows or explicit anisotropic feedback.

When arguing for the SMBH-CGM hypothesis, MN21 point toward two additional results. First, they find that the ASGQ amplitude is stronger for quiescent than for star-forming centrals in SDSS \citep[also see][]{Yang2006_MNRAS_369_1293}. Second, they show that the modulation is stronger for centrals with over-massive SMBHs at fixed halo mass. K07, using detailed mock catalogs, show that the first trend is likely an artifact of group finder errors. Furthermore, it may be attributed to different galaxy--halo alignment strengths between star-forming and quenched galaxies. However, the second finding, that the modulation depends on SMBH mass, is not immediately explained in the K07 scenario. It is reasonable to assume that the mass of the SMBH is correlated with the properties of the central galaxy and host halo, as well as their formation history. This could affect the degree of central--halo alignment and halo shape as a function of SMBH mass. Therefore, even in the scenario proposed by K07, it is reasonable to expect that the the ASGQ amplitude may depend on SMBH mass at fixed halo mass.

Ultimately, we find that the measured ASGQ amplitude in TNG100 can be fully explained by anisotropic subhalo accretion scale factors. Thus, we find no evidence that ASGQ is caused by SMBH-CGM feedback in TNG100, which further cautions against such an interpretation in observations. Interestingly, in IllustrisTNG, we find that the modulation is better predicted from $\aacc$ than from $\mpeak$, as was the case for UniverseMachine. This result is consistent with MN21’s claim that ASGQ in TNG100 is not a product of pre-processing; instead, the ASGQ modulation is caused by satellites quenching within their host halo and being accreted earlier along the central major axis. The slightly different behavior between UniverseMachine and TNG100 may be attributed to how these models treat subhalo and satellite distributions, but this does not change the basic picture of anisotropic accretion properties driving ASGQ.

We also note the importance of survey realism when comparing ASGQ predictions from simulations against observational results. For example, galaxy group finder errors in SDSS have been shown to have the potential to lead to biased or spurious trends inferred from group catalogs, especially those dependent on galaxy color and SFR \citep[see, e.g.][]{Kang2007_MNRAS_378_1531, Campbell2018_MNRAS_477_359, Calderon2018_MNRAS_480_2031}. One should also take into account how stellar masses and SFRs in observations are estimated and potentially biased. Finally, galaxy formation simulations do not perfectly reproduce all obserational results, such as redshift-dependent stellar mass functions and SFR distributions, and this may affect the predicted ASGQ strength. Such a detailed study is beyond the scope of this work but implies that one should not directly compare the ASGQ amplitudes shown in Figures~\ref{fig:um_quenched_mpeak_driver} and \ref{fig:tng_quenched_drivers_2} against SDSS results.

\section{Conclusion}

In this letter, we analyze results from two state-of-the-art galaxy formation simulations, UniverseMachine and IllustrisTNG, in order to evaluate previous claims that three-dimensional SMBH-driven outflows are at the origin of anisotropic satellite galaxy quenching. Confirming earlier results, we find an anisotropic distribution in satellite accretion properties, i.e., that satellites and subhalos along the major axis of the central galaxy vary regarding accretion times and peak halo mass. We argue that in both UniverseMachine and IllustrisTNG, ASGQ can be entirely explained by the anisotropic accretion properties of satellites and basic assumptions about quenching processes. Our results thus caution against the interpretation of anisotropic satellite galaxy quenching as direct observational evidence of both interactions between SMBH outflows and the CGM and their impact on satellite galaxy properties. More generally, our results show that studies exploring spatial anisotropy in satellite populations should consider the possible anisotropy in the subhalos that host them.

\section*{Acknowledgements}

We thank Phil Mansfield, Annalisa Pillepich, Ignacio Mart\'in-Navarro, and Frank C. van den Bosch for stimulating discussions. JK received support from the Yale University Summer Experience Award. This research was also supported by the Kavli Institute for Particle Astrophysics and Cosmology.

We thank the IllustrisTNG collaboration for publicly releasing the TNG100 simulation \citep{Pillepich2018_MNRAS_475_648, Springel2018_MNRAS_475_676, Nelson2018_MNRAS_475_624, Naiman2018_MNRAS_477_1206, Marinacci2018_MNRAS_480_5113, Nelson2019_ComAC_6_2} and Peter Behroozi for making the results of UniverseMachine \citep{Behroozi2019_MNRAS_488_3143} publicly available. This work made use of the following software packages: {\sc Astropy} \citep{AstropyCollaboration2013_AA_558_33}, {\sc halotools} \citep{Hearin2017_AJ_154_190}, {\sc matplotlib} \citep{Hunter2007_CSE_9_90}, {\sc NumPy} \citep{vanderWalt2011_CSE_13_22}, {\sc scikit-learn} \citep{Pedregosa2011_JMLR_12_2825}, and {\sc Spyder}.

The IllustrisTNG simulations were undertaken with computing time awarded by the Gauss Centre for Supercomputing (GCS) under GCS Large-Scale Projects GCS-ILLU and GCS-DWAR on the GCS share of the supercomputer Hazel Hen at the High Performance Computing Center Stuttgart (HLRS), as well as on the machines of the Max Planck Computing and Data Facility (MPCDF) in Garching, Germany. The MultiDark Database used in this paper and the web application providing online access to it were constructed as part of the activities of the German Astrophysical Virtual Observatory as a result of a collaboration between the Leibniz-Institute for Astrophysics Potsdam (AIP) and the Spanish MultiDark Consolider Project CSD2009-00064. The Bolshoi and MultiDark simulations were run on NASA’s Pleiades supercomputer at the NASA Ames Research Center. The MultiDark-Planck (MDPL) and the BigMD simulation suite were performed in the Supermuc supercomputer at LRZ using time granted by PRACE.

\bibliography{bibliography}

\end{document}